\documentclass[aps,prl,twocolumn]{revtex4}
\usepackage{graphicx}
\usepackage{dcolumn}
\usepackage{amsmath}
\usepackage{amssymb}
\usepackage{subfigure,amsmath,verbatim,moreverb,bm}
\def\be{\begin{equation}}
\def\ee{\end{equation}}
\def\ber{\begin{eqnarray}}
\def\eer{\end{eqnarray}}

\def\rv{{\bf r}}

\def\fv{{\bf f}}

\begin{document}
\title{The derivative discontinuity in the strong-interaction limit of density functional theory}
\author{Andr\'e Mirtschink,$^1$ Michael Seidl,$^2$ and Paola Gori-Giorgi$^1$}
\affiliation{$^1$Department of Theoretical Chemistry and Amsterdam Center for Multiscale Modeling, FEW, Vrije Universiteit, De Boelelaan 1083, 1081HV Amsterdam, The Netherlands\\
$^2$Institute of Theoretical Physics, University of Regensburg, D-93040 Regensburg, Germany}
\date{\today}
\begin{abstract}
We generalize the exact strong-interaction limit of the exchange-correlation energy of Kohn-Sham density functional theory to open systems with fluctuating particle numbers. When used in the self-consistent Kohn-Sham procedure on strongly-interacting systems, this functional yields exact features crucial for important applications such as quantum transport. In particular, the step-like structure of the highest-occupied Kohn-Sham eigenvalue is very well captured, with accurate quantitative agreement with exact many-body chemical potentials. Whilst it can be shown that a sharp derivative discontinuity is only present in the infinitely strongly-correlated limit, at finite correlation regimes we observe a slightly-smoothened discontinuity, with qualitative and quantitative features that improve with increasing correlation. From the fundamental point of view, our results  obtain the derivative discontinuity without making the assumptions used in its standard derivation, offering independent support for its existence. 
\end{abstract}
\maketitle
First-principles calculations of many-electron systems such as solids, molecules, and nanostructures are based, to a very large extent, on Kohn-Sham (KS) \cite{KohSha-PR-65} density functional theory (DFT) \cite{HohKoh-PR-64}. KS DFT is, in principle, an exact theory, in which the ground-state energy and density of an interacting many-electron system are mapped into a problem of non-interacting electrons moving in the effective one-body KS potential. In practice, KS DFT relies on approximations for the exchange-correlation energy that, although successful in very many cases, have still deficiencies that hamper its overall usefulness \cite{CohMorYan-CR-12}. 

\begin{figure}
\includegraphics[width=6.0cm]{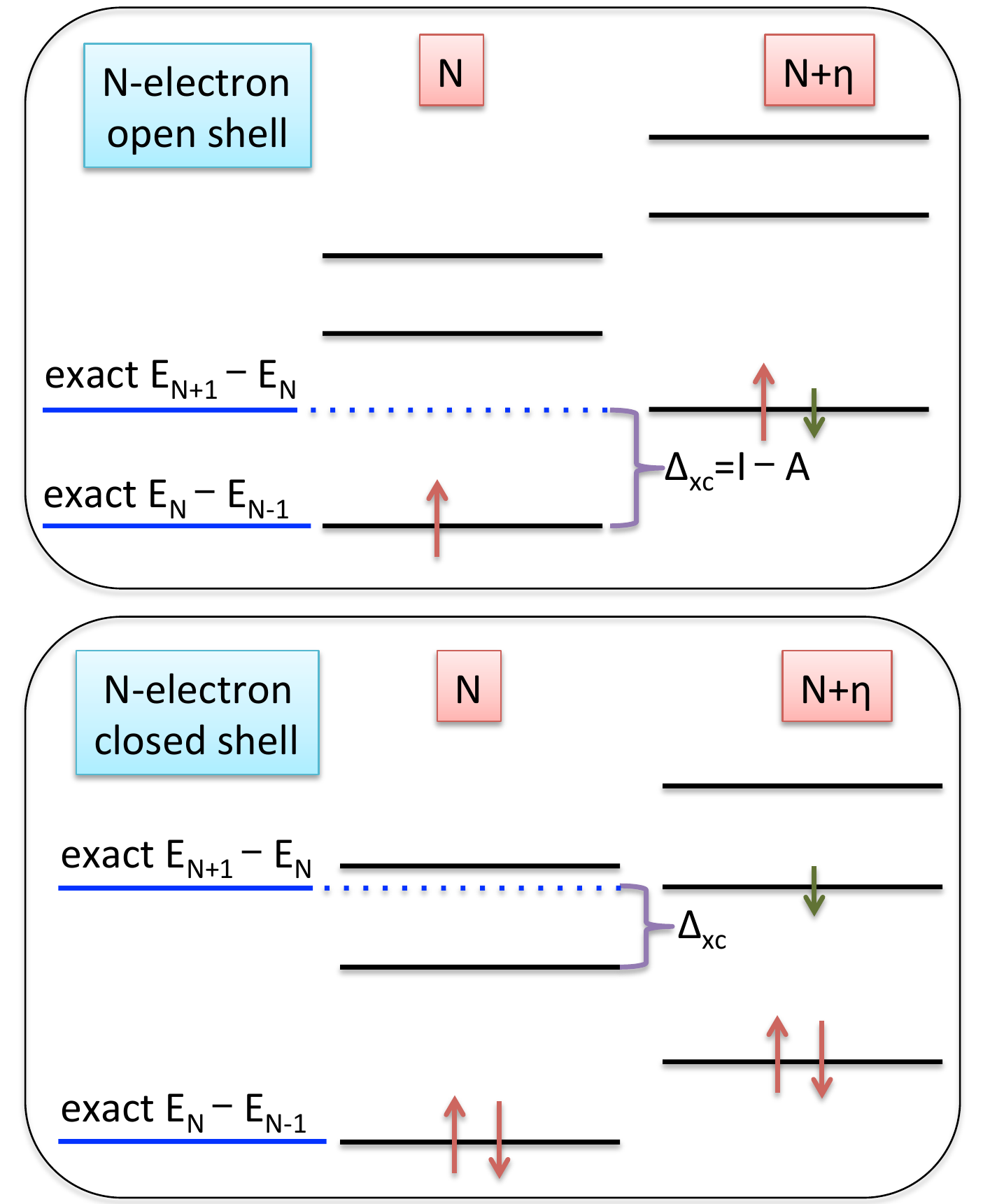}
   \caption{(color online). Schematic illustration of the spin-restricted (same potential for spin-$\uparrow$ and spin-$\downarrow$ electrons) KS spectrum when adding a tiny fraction $\eta$ of an electron to an $N$-electron system. Top panel: when the KS $N$-electron system is open shell, the whole KS spectrum ``jumps'' by a positive constant $\Delta_{xc}=I_N-A_N$, which aligns the KS highest occupied eigenvalue (HOMO) to minus the electron affinity $-A_N=E_{N+1}-E_N$. Bottom panel: when the KS system is closed shell, the positive constant $\Delta_{xc}$ aligns the KS $N$-electron lowest unoccupied orbital (LUMO) to minus the exact affinity $-A_N$. In both cases, in the exact KS system the HOMO is equal to the many-body chemical potential.}
\label{fig_Deltaxc}
\end{figure}

Exact KS DFT has many weird and counterintuitive features often missed by the available approximations. One of the weirdest and most elusive of these features is the derivative discontinuity of the exact exchange-correlation energy functional at integer particle numbers $N$  \cite{PerParLevBal-PRL-82}, which has been an incredibly long-debated issue \cite{GunSch-PRL-86,GodSchSha-PRL-86,Kle-PRB-97,PerLev-PRB-97,Kle-PRBr-97,ZahWan-PRA-04,GruMarRub-PRB-06,SagPer-PRA-08,MorCohYan-PRL-09,GorSav-IJQC-09,BroYinLor-SR-13} because its formal derivation relies on some (very reasonable) assumptions. This discontinuity makes the exact KS potential ``jump'' by a constant $\Delta_{xc}$ when we add to an $N$-electron system even a very tiny fraction $\eta$ of an electron, aligning the chemical potential of the non-interacting KS system to the exact, interacting, one, as schematically illustrated in Fig.~\ref{fig_Deltaxc}.
The derivative discontinuity has crucial physical consequences. For example, it accounts for the difference between the KS and the physical fundamental gaps \cite{PerLev-PRL-83,ShaSch-PRL-83,Per-IJQC-86}, it allows a correct KS DFT description of molecular dissociation \cite{PerParLevBal-PRL-82,CohMorYan-SCI-08}, and it should improve charge-transfer excitations in time-dependent DFT \cite{GriBae-JCP-04,VieCapUll-PCCP-09,HelGro-PRA-12}. It also plays a fundamental role in quantum transport, especially to capture the physics of the Coulomb blockade and the Kondo effect \cite{TohFilSanBur-PRL-05,CapBorKarRei-PRL-07,KurSteKhoVerGro-PRL-10,SteKur-PRL-11,KurSte-PRL-13}.
These are all cases in which the standard approximations, which miss this discontinuity, work poorly. Corrections based on the explicit enforcement of the discontinuity have been often proposed as a practical solution (see, e.g., Refs.~\cite{DabFerPoiLiMarCoc-PRB-10,RefShaGovNeaBaeKro-PRL-12,KroSteRefBae-JCTC-12,KraKro-PRL-13}).

Recently, it has been shown \cite{MalGor-PRL-12,MalMirCreReiGor-PRB-13} that the exact strong-interaction limit of DFT \cite{SeiGorSav-PRA-07,GorVigSei-JCTC-09} provides a non-local density functional, called ``strictly-correlated electrons'' (SCE) \cite{Sei-PRA-99,SeiGorSav-PRA-07,GorVigSei-JCTC-09}, which can be used to approximate the exchange-correlation energy of KS DFT, capturing key features of strong correlation such as charge localization in low-density quantum wires without any symmetry breaking \cite{MalGor-PRL-12,MalMirCreReiGor-PRB-13}. This approximation becomes asymptotically exact in the very low-density (or infinitely strong correlation) limit \cite{MalGor-PRL-12,MalMirCreReiGor-PRB-13}. The purpose of this Letter is to generalize the SCE functional to the case of fractional particle numbers, yielding an exchange-correlation functional for open systems 
that becomes more and more accurate as correlation increases, and that can be already used for modeling nanotransport through low-density quantum wires and dots. Our results also support the assumptions that were made to derive the existence of the derivative discontinuity \cite{PerParLevBal-PRL-82}, and provide new insight for the construction of approximate functionals.

The Letter is organized as follows. After briefly reviewing the essential background material, we present the rigorous generalization of the SCE functional to fractional electron numbers. We then show that, without breaking the spin symmetry, the self-consistent KS results with the SCE functional display the right discontinuities at integer electron numbers when correlation is very strong. This is obtained without making any additional assumption or imposing any {\it ad hoc} constraint. Finally, we draw our conclusions and discuss some perspectives. Hartree (effective) atomic units are used throughout.

{\it Fractional particle numbers in KS DFT--} At zero temperature, open systems with fluctuating particle numbers have been first analyzed in the KS framework by considering  statistical mixtures \cite{PerParLevBal-PRL-82}. When dealing with DFT for quantum mechanical systems, we can work with pure states in which a degenerate system is composed by well separated fragments, and one focuses on the energy and density of one of the fragments alone \cite{YanZhaAye-PRL-00,PerRuzCsoVydScuStaTao-PRA-07,Aye-JMC-08}. A very simple example is a stretched H$_2^+$ molecule, in which on each proton we find, on average, $\frac{1}{2}$ electron \cite{CohMorYan-SCI-08,MorCohYan-PRL-09}. 
A key point to prove the existence of the derivative discontinuity is the (empirical) observation that, for integer particle numbers $N$, the energy $E_N$ of a $N$-electron system in a given external potential is a concave-up function, {\it i.e.} $E_N\le\frac{1}{2}(E_{N+1}+E_{N-1})$. This implies that for fractional electron numbers $Q=N+\eta$ (with $0\le\eta\le 1$) the exact many-electron ground-state energy $E_Q$ and density $\rho_Q(\rv)$ lay on the line connecting the values at the two adjacent integers:
\begin{eqnarray}
E_{N+\eta} & = & (1-\eta)E_N+\eta \,E_{N+1} 
\label{eq_E(N+eta)}\\
\rho_{N+\eta}(\rv) & = & (1-\eta)\rho_N(\rv)+\eta\,\rho_{N+1}(\rv).
\label{eq_rho(N+eta)}
\end{eqnarray}
In KS DFT, one usually aims at obtaining the exact quantities of Eqs.~\eqref{eq_E(N+eta)}-\eqref{eq_rho(N+eta)} by means of a non-interacting system of $Q=N+\eta$ electrons in the effective single-particle KS potential $v_{s}(\rv;[\rho_Q])$, which necessarily changes as we change $Q$. The energy is then minimized by giving integer occupation to the $N$ single particle KS spin-orbitals with lowest eigenvalues and fractional occupation $\eta$ to the frontier orbital(s), often called HOMO (highest occupied molecular orbital) \cite{Jan-PRB-78,VydScuPer-JCP-07}. 

In the exact KS theory, the HOMO eigenvalue $\epsilon_{\rm HOMO}$ is the derivative of the total energy of Eq.~\eqref{eq_E(N+eta)} with respect to the particle number $Q$, $\frac{\partial E_Q}{\partial Q}=\epsilon_{\rm HOMO}$ \cite{Jan-PRB-78,PerParLevBal-PRL-82}. Thus, the exact $\epsilon_{\rm HOMO}$ is constant between any two adjacent integers (say, $N$ and $N+1$) and equal to the interacting chemical potential $E_{N+1}-E_{N}$, jumping to a different value when crossing an integer. This ``step-like'' behavior of the KS $\epsilon_{\rm HOMO}$ is not captured by the standard approximate functionals (see, e.g., Refs.~\cite{VydScuPer-JCP-07,PerRuzCsoVydScuStaTao-PRA-07,CohMorYan-PRB-08}), unless imposed {\it a priori} via additional constraints in a spin-unrestricted framework, as e.g., in Refs.~\cite{DabFerPoiLiMarCoc-PRB-10,RefShaGovNeaBaeKro-PRL-12,KraKro-PRL-13}
The alignment of the exact KS HOMO eigenvalue with the interacting chemical potential implies that the exact KS one-body potential must jump by a constant $\Delta_{xc}$  (the derivative discontinuity) when crossing an integer, $v_{s}(\rv;[\rho_{N^+}])-v_{s}(\rv;[\rho_{N^-}])=\Delta_{xc}$ (see Fig.~\ref{fig_Deltaxc}). 

{\it Strong-interaction limit --}The strong-interaction limit of DFT is given by the SCE functional $V_{ee}^{\rm SCE}[\rho]$, defined as the minimum of the electronic interaction alone over all the wave functions yielding the density $\rho$ \cite{Sei-PRA-99,SeiPerLev-PRA-99,SeiGorSav-PRA-07},
\be
V_{ee}^{\rm SCE}[\rho]=\min_{\Psi\to\rho}\langle\Psi|\hat V_{ee}|\Psi\rangle.
\label{eq_VeeSCE}
\ee
 It can be shown \cite{LevPer-PRA-85,GorSei-PCCP-10,MalGor-PRL-12} that in the low-density (or strong-interaction) limit the exact Hartree and exchange-correlation functional $E_{\rm Hxc}[\rho]$ of KS theory tends asymptotically to $V_{ee}^{\rm SCE}[\rho]$.

Physically, the functional $V_{ee}^{\rm SCE}[\rho]$ portrays the  {\em strict correlation} regime, where the position $\rv$ of one electron determines all the other $N-1$ electronic positions $\rv_i$ through the so-called {\em co-motion functions}, $\rv_i=\fv_i[\rho](\rv)$, some non-local functionals of the density \cite{SeiGorSav-PRA-07,GorVigSei-JCTC-09,GorSeiVig-PRL-09,ButDepGor-PRA-12}. Therefore, the net repulsion on an electron at position ${\bf r}$ due to the other $N-1$ electrons depends on $\rv$ alone and can be {\it exactly} represented \cite{SeiGorSav-PRA-07,GorSeiVig-PRL-09,ButDepGor-PRA-12,MalMirCreReiGor-PRB-13} by a local one-body potential,
\be
\nabla \tilde{v}_{\rm SCE}[\rho](\rv)=-\sum_{i= 2}^N \frac{\rv-\fv_i[\rho](\rv)}{|\rv-\fv_i[\rho](\rv)|^3},
\label{eq_vSCE}
\ee
which is also the functional derivative of $V_{ee}^{\rm SCE}[\rho]$, $\frac{\delta V_{ee}^{\rm SCE}[\rho]}{\delta \rho(\rv)}=\tilde{v}_{\rm SCE}[\rho](\rv)$ \cite{MalGor-PRL-12,MalMirCreReiGor-PRB-13}. In terms of the co-motion functions we have \cite{SeiGorSav-PRA-07}
\be
V_{ee}^{\rm SCE}[\rho]=\sum_{i=1}^{N-1}\sum_{j=i+1}^N
\int d\rv\,\frac{\rho(\rv)}{N}\frac{1}{|\fv_i[\rho](\rv)-\fv_j[\rho](\rv)|}.
\label{eq_VeeSCEformula}
\ee

\begin{figure}
\includegraphics[width=8.cm]{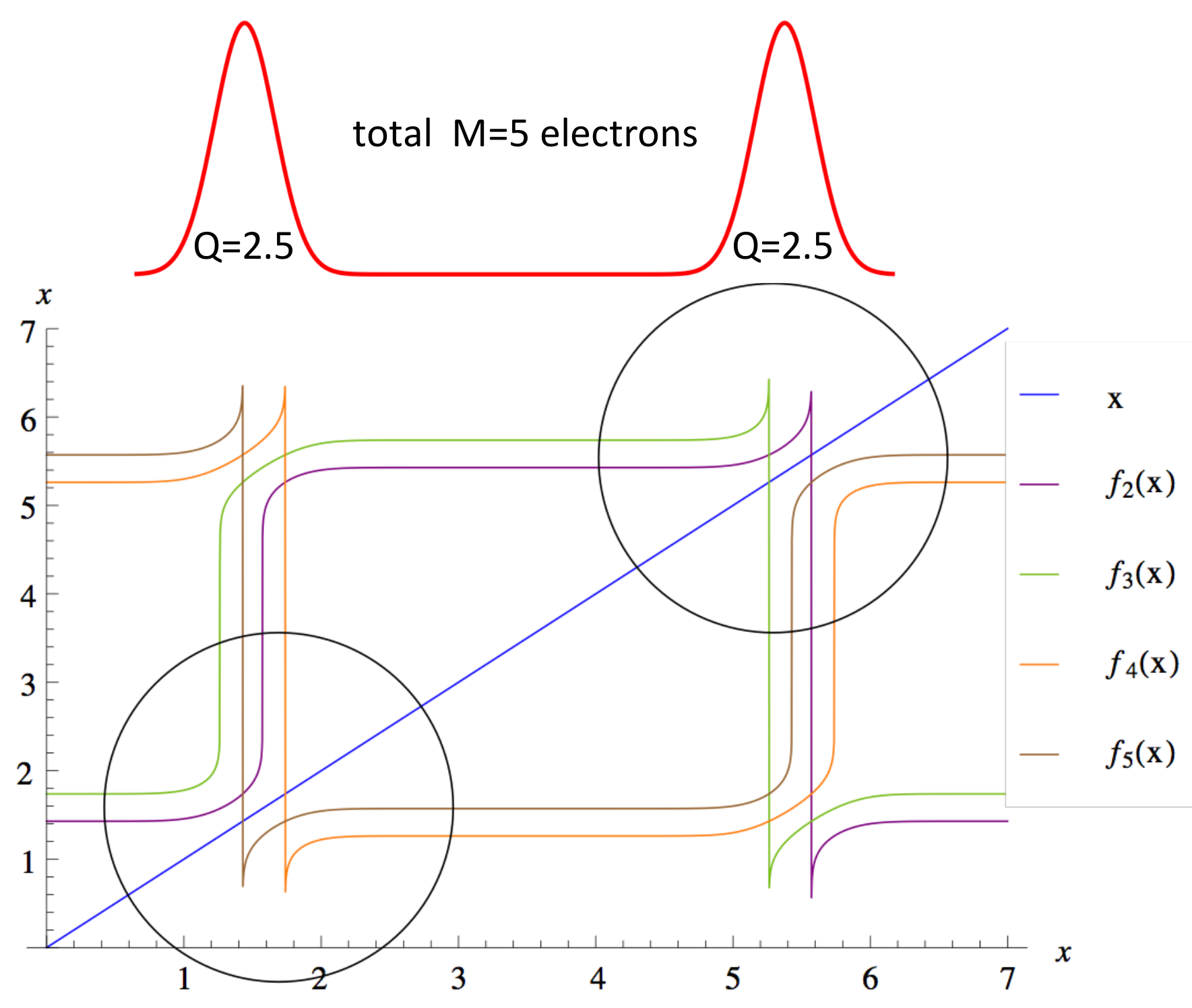}
   \caption{(color online). Simple example of the analysis used to deduce the SCE solution for fractional electron numbers: we considered a density with $M=5$ electrons, made of two separated identical fragments (here modeled with two gaussians), and we have studied the exact SCE solution on each fragment. The graphic shows the positions $f_i(x)$ of the other 4 electrons as a function of the position $x$ of the first electron. The two black circles represent the ``local'' SCE solution on each fragment.}
\label{fig_fragments}
\end{figure}

{\it SCE for fractional particle numbers --} The generalization of the SCE formalism to non-integer particle numbers $Q$ is not obvious, because in Eqs.~\eqref{eq_vSCE}-\eqref{eq_VeeSCEformula} the sum runs over the integer number of electrons $N$. To proceed in a rigorous way \cite{YanZhaAye-PRL-00,Aye-JMC-08,CohMorYan-SCI-08,MorCohYan-PRL-09,CohMorYan-CR-12},  we first analyze a well separated $Q$-electron fragment inside a degenerate system with total integer electron number $M$, as schematically illustrated in Fig.~\ref{fig_fragments}.

Actually, from Eq.~\eqref{eq_VeeSCEformula} it is not evident that $V_{ee}^{\rm SCE}[\rho]$ separates into the  sum of contributions of the isolated fragments, because the interaction $|\fv_i[\rho](\rv)-\fv_j[\rho](\rv)|^{-1}$ between two electrons on a given fragment may contribute significantly to the integral when $\rv$ spans the region of another fragment. However, we have recently shown \cite{MirSeiGor-JCTC-12} that another exact expression for $V_{ee}^{\rm SCE}[\rho]$ is
\be
V_{ee}^{\rm SCE}[\rho]=\frac{1}{2}\int d\rv\,\rho(\rv)\sum_{k=2}^N\frac{1}{|\rv-\fv_k[\rho](\rv)|},
\label{eq_VeeSCEsizeconsist}
\ee 
which separates into a sum of fragment contributions, showing explicitly that $V_{ee}^{\rm SCE}[\rho]$ is size consistent.

We start from (quasi) one-dimensional systems (1D), for which we can construct the SCE solution analytically. The main findings and conclusions, however, should hold also for two- and three-dimensional (3D) systems \cite{WagStoBurWhi-PCCP-12}. Indeed, we have also performed a three-dimensional self-consistent calculation for a spherically-symmetric density, for which we can deduce the fractional SCE solution from our 1D construction, finding similar results.

Figure \ref{fig_fragments} depicts the reasoning behind the construction of the SCE functional for $Q$ electrons. In the illustrated example, we have solved the SCE problem for $M=5$ electrons, for a density made of two well-separated identical fragments, each integrating to $Q=2.5=N+0.5$ particles. We have then studied the ``local'' SCE solution on the fragments, indicated in  Fig.~\ref{fig_fragments} by  black circles. Here, the positions $f_i(x)$ of the other 4 electrons is shown as a function of the position $x$ of the first electron. We find that two adjacent strictly-correlated positions $f_i(x)$ and $f_{i+1}(x)$ on the fragment always satisfy the condition of total suppression of fluctuations \cite{Sei-PRA-99},
\begin{equation}
	\int_{f_i(x)}^{f_{i+1}(x)}\rho(x')dx'=1,
\label{eq_supprfluct}
\end{equation}
so that there are values of $x$ for which we have 3 electrons in the fragment, and values of $x$ for which we find only 2 particles (the third electron is in the other fragment).

Similarly, we find that the general SCE solution for a density integrating to $Q=N+\eta$ electrons can be easily obtained from Eq.~\eqref{eq_supprfluct} (see the Supplemental Material for a detailed derivation). From now on, we work with the fragment alone, considering that on our $x$-axis only the density $\rho_Q(x)$ is present. The co-motion functions read
\begin{align}
	f_i(x)=	\begin{cases}
				N_e^{-1}[N_e(x)+i-1] 	& x<a_{N+1+\eta-i}\\
				N_e^{-1}[N_e(x)+i-N-2] 	& x>a_{N+2-i}\\
				\infty			&\text{otherwise,}
			\end{cases}
			\label{eq:comf}
\end{align}	
where the function $N_e(x)$ is defined via the density $\rho_Q(x)$,
\begin{equation}
	N_e(x)=\int_{-\infty}^x \rho_Q(x')dx',
	\label{eq_Ne}
\end{equation}
$a_k=N_e^{-1}(k)$, and $i=2,...,N+1$. Notice that even if we have $N+1$ co-motion functions, when  $x\in [a_{N+1+\eta-i},a_{N+2-i}]$ one of the electrons stays at infinity, so that there are $x$-intervals for which we find $N+1$ electrons in the density, and $x$-intervals for which we have only $N$ electrons (one of the electrons cannot ``enter'' in the density).

{\it Self-consistent KS SCE for $N+\eta$ electrons --} From Eqs.~\eqref{eq:comf}, \eqref{eq_VeeSCEsizeconsist} and \eqref{eq_vSCE} we can now construct the SCE functional $V_{ee}^{\rm SCE}[\rho_Q]$ and its functional derivative $\tilde{v}_{\rm SCE}[\rho_Q](\rv)$ for any one-dimensional density integrating to a non-integer particle number $Q$. [For $Q=N+\eta$ electrons, the sum in Eqs.~\eqref{eq_VeeSCEsizeconsist} and \eqref{eq_vSCE} runs up to $N+1$.]
We then consider $Q$ electrons in the quasi-one-dimensional model quantum wire of 
Refs.~\cite{BedSzaChwAda-PRB-03,AbePolXiaTos-EJPB-07}, in which the effective 
electron-electron interaction is obtained by integrating the Coulomb repulsion on the lateral 
degrees of freedom \cite{CalGol-PRB-97,BedSzaChwAda-PRB-03}, and is given by 
$w_b(x)=\frac{\sqrt{\pi}}{2\,b}\,\exp\left(\frac{x^2}{4\,b^2}\right){\rm erfc}\left(\frac{x}{2\,b}\right)$.
The parameter $b$ fixes the thickness of the wire, set to $b=0.1$ throughout this work, and ${\rm erfc}(x)$ is the complementary error function. As in Ref.~\cite{AbePolXiaTos-EJPB-07}, we consider an external harmonic confinement $v_{\rm ext}(x)=\frac{1}{2}\omega^2x^2$ in the direction of motion of the electrons, which
for small $\omega$ yields bound states with very low density, where the functional $V_{ee}^{\rm SCE}[\rho]$ becomes closer and closer to the exact $E_{\rm Hxc}[\rho]$. This is crucial to show that when $E_{\rm Hxc}[\rho]\to V_{ee}^{\rm SCE}[\rho]$ we recover the exact features of Fig.~\ref{fig_Deltaxc}.

We have then performed self-consistent, spin-restricted (same KS potential for $\uparrow$ and $\downarrow$ spins), KS calculations with the SCE potential, as in Refs.~\cite{MalGor-PRL-12,MalMirCreReiGor-PRB-13}, this time by giving fractional occupation to the KS HOMO orbital \cite{Jan-PRB-78,VydScuPer-JCP-07,GaiFirSta-PRL-12}. The SCE potential $\tilde{v}_{\rm SCE}[\rho_Q](x)$, representing the approximate Hartree plus exchange-correlation potentials in our KS SCE calculations, is obtained by integrating the 1D analogue of Eq.~\eqref{eq_vSCE}, with boundary condition $\tilde{v}_{\rm SCE}[\rho_Q](|x|\to\infty)=0$, using the co-motion functions of Eq.~\eqref{eq:comf}. 
\begin{figure}
\includegraphics[width=6.5cm]{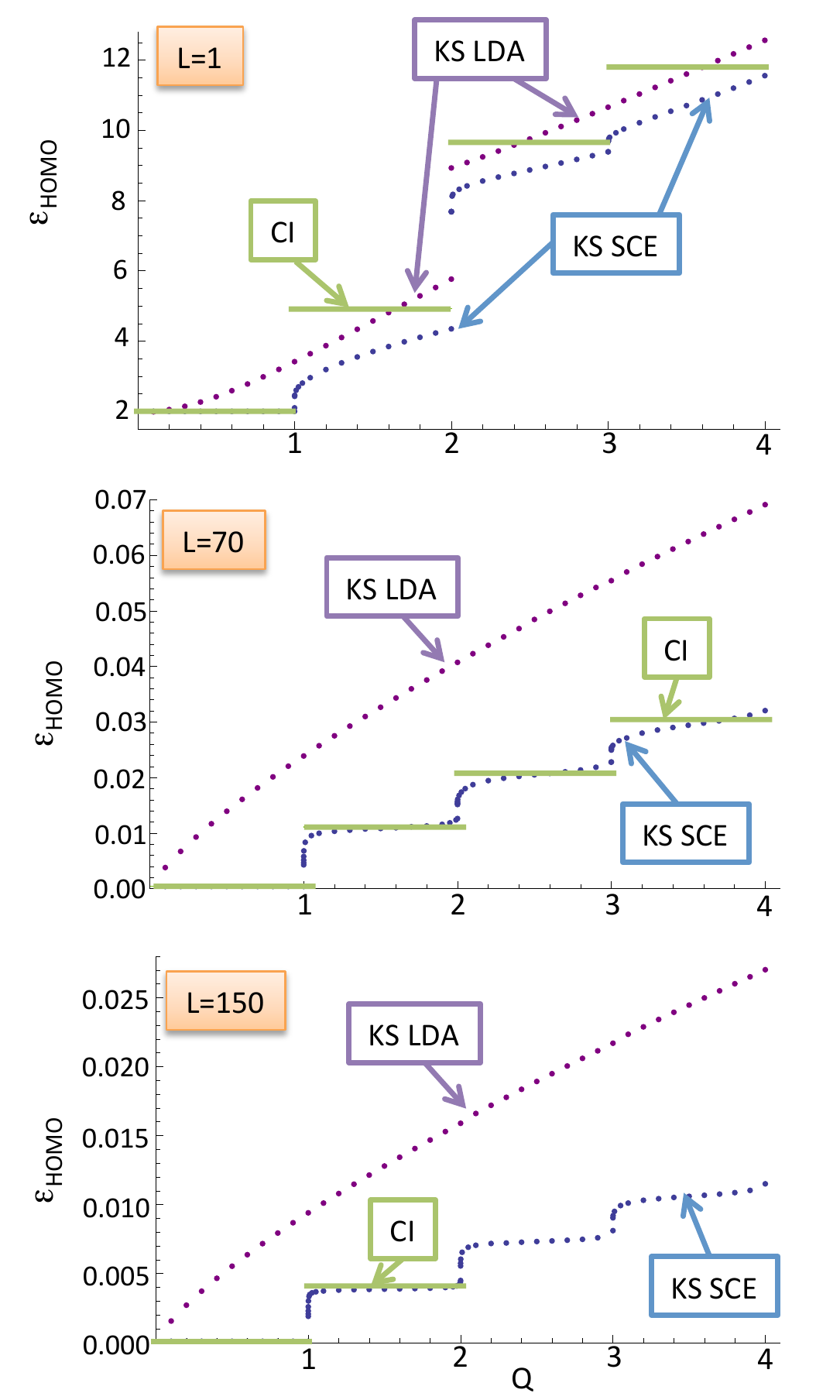}
   \caption{(color online). Self-consistent results for the spin-restricted KS HOMO eigenvalue (in Hartrees) as a function of the particle number $Q$ for a quasi-1D quantum wire with harmonic confinement along the direction of motion of the electrons, $v_{\rm ext}(x)=\frac{1}{2}\omega^2x^2$, and $\omega=\frac{4}{L^2}$. The KS SCE results are compared with the standard KS LDA and with the exact chemical potential $E_{N+1}-E_N$ from full CI calculations \cite{MalMirCreReiGor-PRB-13}.}
\label{fig_HOMO1D}
\end{figure}

In Fig.~\ref{fig_HOMO1D} we show our self-consistent KS SCE results for the HOMO eigenvalue as a function of the particle number, comparing with the KS LDA values (the 1D LDA functional is from Ref.~\cite{CasSorSen-PRB-06}), and with the full configuration-interaction (CI) results \cite{MalMirCreReiGor-PRB-13} for the chemical potential $E_{N+1}-E_N$. We have considered a moderately-correlated case, $L=1$, and two strongly-correlated cases, $L=70$ and $L=150$, where $L$ is an effective confinement-length such that $\omega=\frac{4}{L^2}$. For moderate correlation, $L=1$, we see that LDA, having $\Delta_{xc}=0$, shows a discontinuity in the HOMO value only when filling a new shell (even $N$; case illustrated in the second panel of Fig.~\ref{fig_Deltaxc}), while KS SCE shows a small vertical change also when the $N$ system is open shell (at odd $N$). When correlation becomes stronger ($L=70$ and 150), the KS SCE self-consistent HOMO approaches more and more the exact step structure, with very good quantitative agreement with the full CI chemical potentials. For such cases, KS LDA yields essentially a continuous curve, since the single particle energies are all very close. 

\begin{figure}
\includegraphics[width=6.7cm]{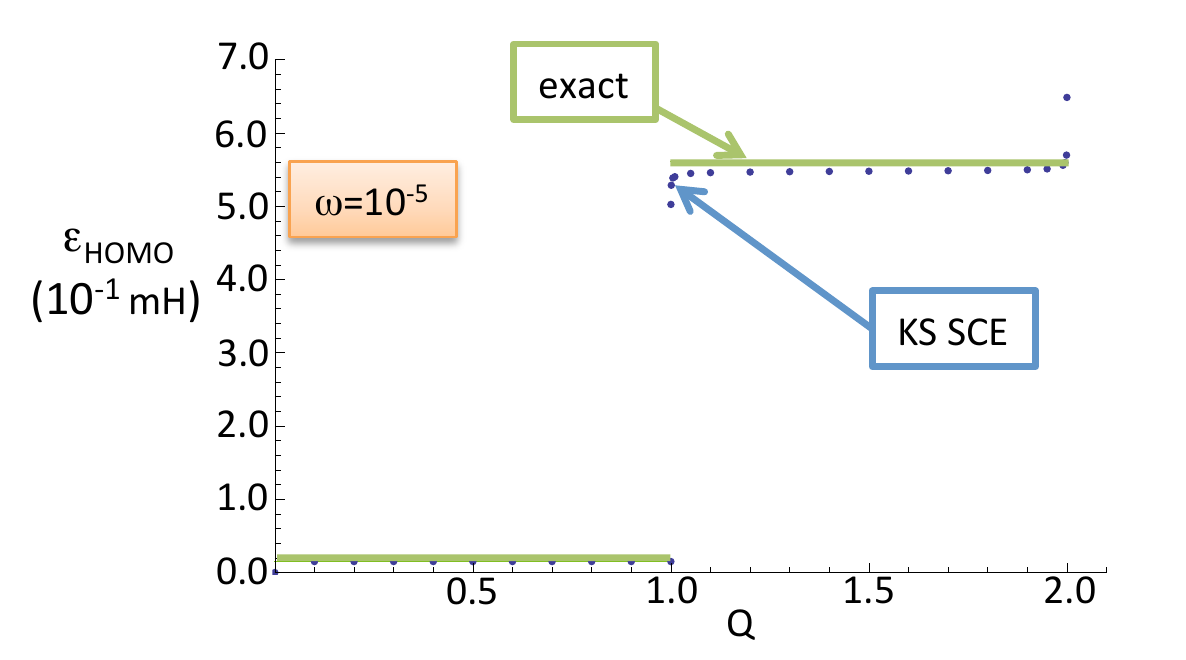}
   \caption{(color online). Self-consistent results for the spin-restricted KS HOMO eigenvalue as a function of the particle number $Q$ for a three-dimensional electronic system in the external harmonic potential, $v_{\rm ext}(\rv)=\frac{1}{2}\omega^2r^2$, with $\omega=10^{-5}$. The KS SCE results are compared with the exact chemical potential $E_{N+1}-E_N$ \cite{Tau-PRA-93,CioPer-JCP-00}.}
\label{fig_HOMO3D}
\end{figure}

In Fig.~\ref{fig_HOMO3D}, we show the result for a 3D system, in which the electron-electron interaction is Coulombic and the external potential is harmonic (``Hooke's atom''). The co-motion functions can be deduced from Eq.~\eqref{eq:comf}, as detailed in the Supplemental Material. For small $\omega$, we find that the self-consistent KS SCE HOMO approaches again the exact step structure, which becomes sharper and sharper as correlation increases ($\omega$ decreases) \footnote{For the failure of standard functionals at this correlation regime see, e.g., M. Taut, A. Ernst, and H. Eschrig, J. Phys. B: At. Mol. Opt. Phys. {\bf 31}, 2689 (1998).}, similarly to the 1D results of Fig.~\ref{fig_HOMO1D}. 
Notice, again, that for $0\le Q\le 2$, we have one KS orbital with occupancy $Q$.

{\it Concluding remarks and perspectives--} 
The discontinuity in the HOMO for open shell systems (case in the upper panel of Fig.~\ref{fig_Deltaxc}) from the self-consistent KS SCE is not just a unique result, but also an independent proof that the exact KS formalism should have this feature. In fact, we have only used the exact $V_{ee}^{\rm SCE}[\rho_Q]$ in the KS self-consistent procedure, without imposing any other condition on our functional. Until now, this feature has only been captured in the context of lattice hamiltonians \cite{LimSilOliCap-PRL-03}, or by imposing it {\it a priori} in a spin-unrestricted framework \footnote{An exception to the spin-unrestricted construction is Ref.~\cite{MorCohYan-PRL-09}, which attempts at constructing, in a spin-restricted framework,  a functional showing the linear behavior of Eq.~\eqref{eq_E(N+eta)}.}, as, e.g., in Refs.~\cite{DabFerPoiLiMarCoc-PRB-10,RefShaGovNeaBaeKro-PRL-12,KraKro-PRL-13}. 

From a practical point of view, our results could already be used to model transport through a correlated quantum wire or quantum dot, going beyond the lattice calculations of Refs.~\cite{KurSteKhoVerGro-PRL-10,KurSte-PRL-13}. Our findings also provide novel insight for the construction of approximate exchange-correlation functionals: the challenge is to transfer this exact behavior into approximations for less extreme correlation regimes relevant for solid-state physics and chemistry (see Supplemental Material for a test on the H atom). Another challenge is the construction of the SCE functional for general 3D geometry, for which a promising route (that can be generalized to open systems along the lines of this work) is given in Refs.~\onlinecite{ButDepGor-PRA-12,MenLin-PRB-13}.

We thank E. J. Baerends, O. Gritsenko, F. Malet, V. Staroverov and G. Vignale for many insightful discussions. This work was supported by the Netherlands Organization for Scientific Research (NWO) through a Vidi grant.

\end{document}